\documentclass[12pt,showpacs,]{iopart}
\usepackage{graphicx, epsfig}
\begin{document}
\title{Anomalous Magnetic Properties in Ni$_{50}$Mn$_{35}$In$_{15}$}
\author{P. A. Bhobe\dag\footnote[3]{To whom correspondence should be addressed
(preeti@tifr.res.in)}, K. R. Priolkar\ddag and A. K. Nigam\dag}

\address{\dag\ Tata Institute of Fundamental Research, Homi Bhabha
Road, Mumbai 400 005 India}

\address{\ddag\ Department of Physics, Goa University, Goa, 403 206 India}

\date{\today}

\begin{abstract}
We present here a comprehensive investigation of the magnetic
ordering in Ni$_{50}$Mn$_{35}$In$_{15}$ composition. A concomitant
first order martensitic transition and the magnetic ordering
occurring in this off-stoichiometric Heusler compound at room
temperature signifies the multifunctional character of this magnetic
shape memory alloy. Unusual features are observed in the dependence
of the magnetization on temperature that can be ascribed to a
frustrated magnetic order. It is compelling to ascribe these
features to the cluster type description that may arise due to
inhomogeneity in the distribution of magnetic atoms. However,
evidences are presented from our ac susceptibility, electrical
resistivity and dc magnetization studies that there exists a
competing ferromagnetic and antiferromagnetic order within crystal
structure of this system. We show that excess Mn atoms that
substitute the In atoms have a crucial bearing on the magnetic order
of this compound. These excess Mn atoms are antiferromagnetically
aligned to the other Mn, which explains the peculiar dependence of
magnetization on temperature.
\end{abstract}

\submitto{\JPD} 
\maketitle

\section{Introduction}
Ni-Mn based Heusler alloys of the type Ni$_{50}$Mn$_{50-y}$X$_y$ ( X
= In, Sn and Sb) have recently been identified by Sotou {\it et.al}
\cite{suto} as systems undergoing martensitic transition in the
ferromagnetic state. Such compounds have the tendency to display
huge strains with application of moderate magnetic fields. This
unique magnetoelastic property has numerous technological
implications leading to wide spread applied research like the
magnetic shape memory, magnetocaloric and magnetoresisitive effect.
These unique properties are an outcome of the {\it Martensitic Phase
Transformation} that takes place in a magnetically ordered state.
The classic example of this class of materials is the Heusler alloy,
Ni$_2$MnGa \cite{web, ulla, tick, sozi}.

Martensitic transformation is a first-order structural phase change
wherein the constituent atoms get displaced from its
crystallographic positions over varied amplitude of displacement in
a highly correlated fashion. Upon transformation, the crystal
structure changes from a highly symmetric cubic phase to a low
symmetry structure. Some ferromagnetic Heusler compounds are known
to undergo a martensitic transition that is highly reversible and
the compounds can be cycled several times through the transformation
temperature. The generic formula for Heusler compounds is X$_2$YZ.
In the high temperature phase, the structure can be viewed as four
interpenetrating fcc lattices with X atoms occupying the (0,0,0) and
($1\over2$,0,0) sites; Y atoms occupying the
($1\over4$,$1\over4$,$1\over4$) sites and Z atoms occupying the
($3\over4$,$1\over4$,$1\over4$) sites. A tetragonal or orthorhombic
structure is observed in the low temperature martensitic phase with
the displacement of atoms giving rise to modulations that extend
over several crystal planes.

Recently found Ni$_{50}$Mn$_{50-x}$In$_{x}$ compounds that belong to
the class of Heusler alloys have been in focus, especially for $x$
having martensitic transition temperatures near room temperature. In
particular, the stoichiometric composition
Ni$_{50}$Mn$_{25}$In$_{25}$ is known to order ferromagnetically at
T$_C$ $\sim$ 315 K \cite{web-book}, does not undergo a martensitic
transition. While, the Mn-rich off-stoichiometric composition
Ni$_{50}$Mn$_{50-x}$In$_{x}$ exemplifies a martensitic transition.
This transition temperature (T$_M$) is highly sensitive to the Mn:In
ratio and a value between 260 K to 302 K have been reported for In
content varying over a small value of 13 to 16\% atomic
concentration \cite{suto, acet}. However, the T$_C$ is not very
influenced by the Mn:In ratio and varies only slightly with the
composition. A large negative magnetoresistance over 50\% is
attainable in the Ni-Mn-In compound at moderate field strengths
\cite{yu}. A giant isothermal entropy change takes place when the
structural and the magnetic transition temperatures coincide or are
in close vicinity to each other resulting in an inverse
magnetocaloric effect (MCE) near room temperature \cite{ali1,
pab-apl,roy}. It may be noted that an {\it inverse} MCE is generally
displayed by materials with antiferromagnetic order and the
magnitude obtained is generally quite low. It is therefore
fundamentally important to gain thorough insight into the structural
and magnetic aspects of the technologically important compound. With
this aim, we propose to study the nature of magnetic interactions in
the composition Ni$_{50}$Mn$_{35}$In$_{15}$ where a concomitant
structural and magnetic transition is obtained near room
temperature. The key question to be addressed in the present study
is the character of the various magnetic interactions that develop
with varying distances between magnetic atoms in the two
crystallographic phases. We have carried out the measurement of
magnetic and transport properties of Ni$_{50}$Mn$_{35}$In$_{15}$
Heusler alloy and the results are presented here.

\section{Experimental}
Ni$_{50}$Mn$_{35}$In$_{15}$ was prepared by arc-melting the
constituent elements of 4N purity under argon atmosphere. The button
so obtained, was annealed at 1000 K for 48 hours followed by
quenching in ice water. Subsequent energy dispersive x-ray (EDX)
analysis confirmed the composition to be close to nominal with Ni =
49.26, Mn = 35.13 and In = 15.49. The ac susceptibility
($\chi_{ac}$) and four probe resistivity measurements were performed
using Quantum Design Physical Properties Measurement System. While
for the dc magnetization measurements the Quantum Design SQUID
magnetometers (MPMS-XL and MPMS SQUID-VSM) were employed. The
$\chi_{ac}$ Vs. temperature was measured in the presence of
different dc fields and with the excitation frequency varied over
three decades. The sample was cooled to lowest measurement
temperature (5 K) in zero field state and the data was recorded
while warming up to $\sim$ 330 K. Magnetization as a function of
field was measured under sweep magnetic fields up to ${\pm}$ 5 T at
various temperatures. Before each measurement, the sample was heated
to 330 K and cooled in zero field state to the desired temperature.
For the resistivity measurements the data was recorded in the region
5 K to 380 K in the presence of 5 T and 9 T magnetic fields.

\section{Results and Discussion}
\begin{figure}
\epsfig{file=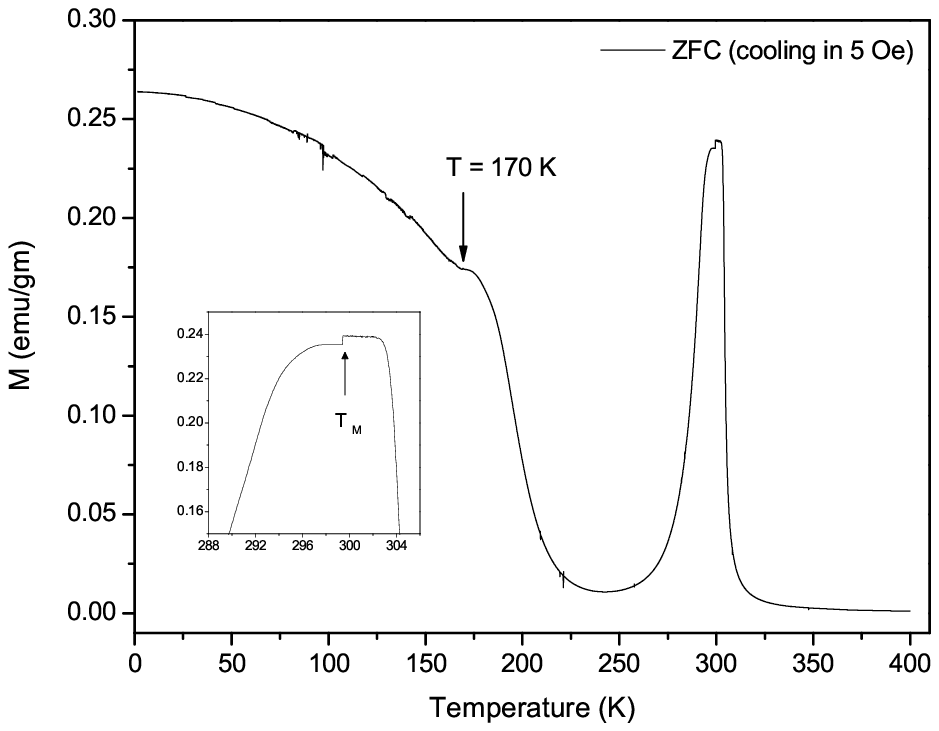, width=8cm, height=7cm}
\centering\caption{\label{vsm-zfc}Magnetization as a function of
temperature measured while cooling in zero field state with nominal
field of 5 Oe. Inset shows the drop in magnetization that occurs
upon martensitic transition.}
\end{figure}

\begin{figure}
\epsfig{file=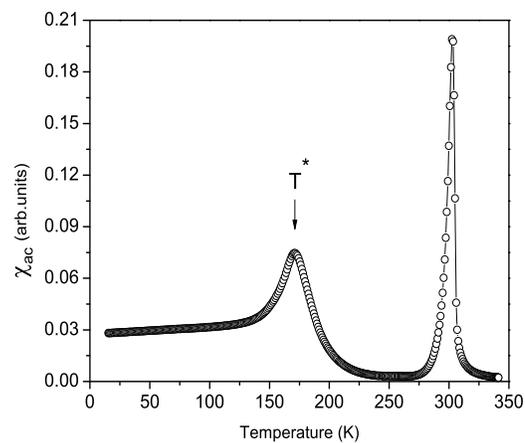, width=8cm, height=7cm}
\centering\caption{\label{acchi}The real component of $\chi_{ac}$
for Ni$_{50}$Mn$_{35}$In$_{15}$ measured after cooling the the
sample from room temperature to 5 K in zero field.}
\end{figure}

The temperature dependence of the magnetization from 5 to 380 K,
recorded while cooling in the zero field state (nominal field 5 Oe),
for Ni$_{50}$Mn$_{35}$In$_{15}$ is presented in figure
\ref{vsm-zfc}. It can be clearly seen that with decreasing
temperature the magnetization rises abruptly at the ferromagnetic
ordering temperature T$_C$ = 306 K. While the abrupt decrease in
magnetization occurs at the martensitic phase transformation with
T$_M$ = 302 K. The inset shows the enlarged view of these magnetic
and martensitic transformations. $\chi_{ac}$ measured at a frequency
{\it f} = 13 Hz in an ac field of 1 Oe is presented in figure
\ref{acchi}. The data shows a very sharp peak at about 300 K which
is an outcome of concomitant martensitic and magnetic transformation
taking place in the compound.

As mentioned earlier, the martensitic transformation is a first
order structural transformation, taking place from a high symmetry
cubic phase to a low-symmetry structure. It thus involves a start
temperature, when the structure starts deforming and a finish
temperature, when the transformation is complete. The variants of
the new crystallographic phase that are formed in this region of the
start and finish temperatures, re-establish the magnetic
interactions that had been present in the parent cubic phase. The
difference in anisotropy strongly modifies the field dependence of
the magnetization in these two phases \cite{albe}. For the present
sample, the structure starts deforming at 302 K and the
transformation completes by $\sim$ 270 K. However, it is interesting
to note that magnetization or $\chi_{ac}$ attains an almost zero
value after the structural transition is complete. This is an
uncommon feature and implies that the magnetic interactions of this
compound should be quite complex.

\begin{figure}
\centering\epsfig{file=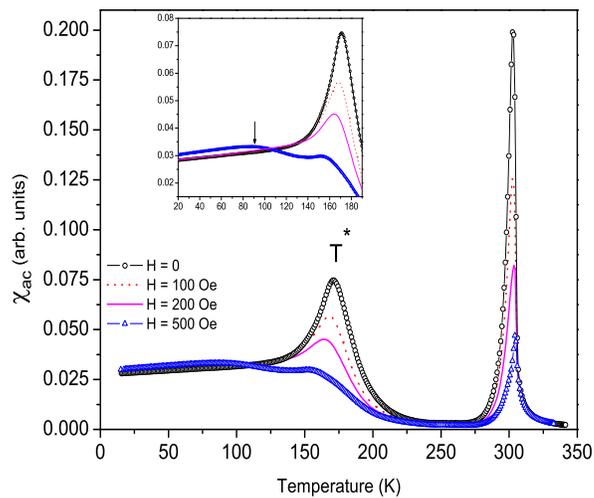, width=9cm, height=8cm}
\caption{\label{chi-field}(Colour online) $\chi_{ac}$(T) measured at
{\it f} = 133 Hz in different dc fields. The intensity of the
peak-like anomaly at T$^*$ decreases with increasing field. The
inset shows the appearance of a hump like feature at a lower
temperature in the H = 500 Oe data after the peak at T$^*$ has
considerably been suppressed.}
\end{figure}

The $\chi_{ac}$ measured during the warm-up cycle displays yet
another broad peak at T$^*$ = 170 K, while the dc magnetization
(measured while cooling) shows a constantly increasing magnetization
with a hump-like feature at the same temperature. However, the
magnetization keeps increasing with the subsequent fall in
temperature below T$^*$. The exact nature of T$^*$ is not clear at
the moment. The first report on this family of compounds by
reference \cite{suto} claims the occurrence of an additional
martensite-to-martensite transformation at low temperatures. They
observe some intricate changes similar to T$^*$ in their low
temperature magnetization data (M(T) with H = 500 Oe) that are taken
as signatures for the occurrence of second structural
transformation. However, feature similar to T$^*$ observed in the
magnetization measurements by reference \cite{acet} has been
ascribed to the magnetic ordering of the martensitic phase by them.
Thus, it becomes essential to verify the nature of T$^*$ and also to
investigate the reason for the drastic fall in $\chi_{ac}$ to almost
zero value upon martensitic transformation. Hence $\chi_{ac}(T)$ was
measured in a constant dc field varying from 0 to 500 Oe and the
plots are presented in figure \ref{chi-field}. With the increase in
dc field the peak at T$^*$ broadens and decreases in magnitude. At
500 Oe, the peak smears out completely and a small hump begins to
appear at a lower temperature, as can be seen in the inset of figure
\ref{chi-field}. Such dependence of the peak at T$^*$ on dc magnetic
field implies that there exists other competing magnetic
interactions in this system that compete with the long range
ferromagnetic order. Such complicated behaviour with competing
ferromagnetic (FM) and antiferromagnetic (AFM) interactions
generally results in a frustrated magnetic order.

\begin{figure}
\epsfig{file=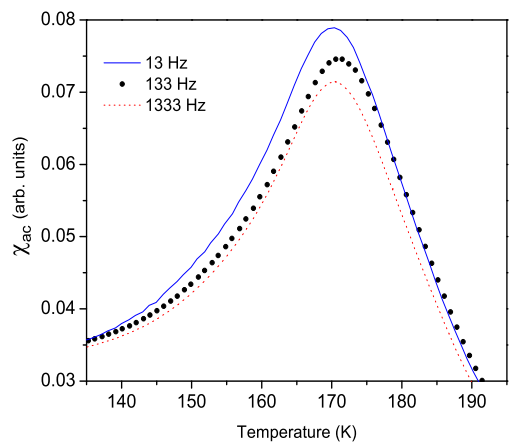, width=8cm, height=7cm}
\centering\caption{\label{ac-freq}(Colour online) The region around
T$^*$ is shown for the real component of $\chi_{ac}$ vs T measured
at different frequencies. T$^*$ does not show any systematic
dependence on {\it f}.}
\end{figure}

A possibility that may give rise to competing FM/ AFM interactions
in Ni$_{50}$Mn$_{35}$In$_{15}$ is the inhomogeneous distribution of
the constituent elements forming regions of varied stoichiometries.
Since the amount of In is the least in this compound, regions rich
in Ni$_2$MnIn and NiMn can form. It is well established that
Ni$_2$MnIn is ferromagnetic in nature \cite{web} while NiMn displays
antiferromagnetism \cite{kas}. In case such a possibility exists in
the present compound, the random distribution of such magnetic
entities would lead to cluster formation and eventually result in
the freezing of the resultant magnetic moment. It is important
mention here that the present compound shows robust martensitic
phase transformation at room temperature while neither Ni$_2$MnIn or
NiMn display this property. Hence little doubt exists about
Ni$_{50}$Mn$_{35}$In$_{15}$ lacking in compositional uniformity.
Nonetheless, to further investigate the cause for FM/AFM
interactions due to possibility of inhomogeneous mixing, $\chi_{ac}$
was measured with varying frequencies. Temperature dependence of
$\chi_{ac}$ at different frequencies {\it f} (13 to 1333 Hz) is
presented at figure \ref{ac-freq}. It is evident from this plots
that the peak position at T$^*$ does not show a shift in temperature
with changing frequencies. This observation rules out the
possibility of T$^*$ being a time dependent phenomenon and hence
cannot be related to freezing of the magnetic moment like in a
spin-glass state.

\begin{figure}
\centering \epsfig{file=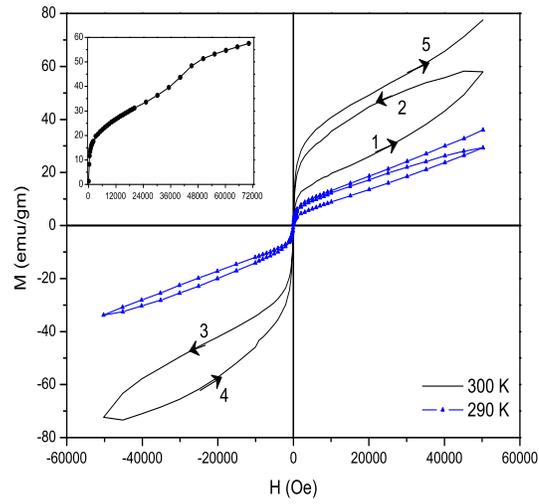, width=8cm, height=8cm}
\caption{\label{mart-loop}(Colour online) Magnetization as a
function of applied field recorded in the vicinity of the
martensitic transition. The metamagnetic character is seen at
intermediate field values as shown in the inset.}
\end{figure}

The shape of the hysteresis loop generally provides a better
understanding of the magnetic ground state of a material. Hence M(H)
loops at select temperatures were obtained by heating the sample to
the paramagnetic region (350 K) and cooling to the desired
temperature in a zero field state. As can be witnessed from figure
\ref{mart-loop}, the M(H) loop at 300 K (and 290 K) displays a very
complex behaviour. The initial steep rise in M(H) for small field
values demonstrates the ferromagnetic character of the sample. With
the increase in field to intermediate values, a metamagnetic
transition is observed as depicted in the inset. In the case of
ferromagnetic shape memory alloys, M(H) measurements in the
martensitic transformation region does display a metamagnetic
transition due to reorientation of the already formed martensite
variants. However, the type of metamagnetism seen in the present
case is quite unusual. Also, there is a considerable increase in the
saturation magnetization upon cycling the sample though magnetic
field. The mechanism driving the transition between the two
metastable states cannot be explained by a simple reorientation of
an existing martensite component, but may be connected to the
nucleation of an additional magnetic phase. It is rather difficult
to attribute the exact cause for the metamagnetic transition.
Further, the width of the hysteresis almost goes to zero at H = 0.
Such hysteresis have been considered earlier as possible signatures
of a field-induced firstorder metamagnetic transition from AFM to FM
in cubic laves phase Co-doped CeFe$_2$ alloys \cite{N-ali}. Also,
with the drop in temperature, there is a decrease in the overall
saturation magnetization seen for the M(H) at 290 K. This
observation is in agreement with the $\chi(T)$ plot where the
susceptibility decreases with decreasing temperature in this
temperature region and starts building up below 240 K with a peak at
T$^*$. Thus M(H) was measured in the vicinity of T$^*$.

Initially the magnetization rises sharply at small field values
indicating a ferromagnetic order. However, it does not saturate at
high fields as expected for a typical ferromagnet. Also, interesting
features are seen in the low field region of M(H) as displayed in
figure \ref{all-loop}. Small hysteresis is observed at 170 K and 160
K. For the M(H) at 100 K and below, the virgin curve initially shows
a linear rise in magnetization with field up to $\sim$ 500 Oe and
then the slope of the curve changes and it lies outside the
hysteresis loop. Such a feature resembles to that of a field-induced
transition. We can then define H$_{cric} \sim$ 500 Oe as a crossover
field. This value of crossover field remains roughly the same for
all the subsequent M(H) curves. It is coincidental that this value
of crossover field is the same as the dc field applied in the
$\chi_{ac}$(T) measurements at which the peak at T$^*$ smears out
completely. The overall features observed in the M(H) curves clearly
demonstrate that the Ni-Mn-In system does not show a pure
ferromagnetic order. Hence $T^*$ cannot be assigned to be the FM
ordering temperature of the martensitic phase. The competing
magnetic interactions present in the system do not allow the
ferromagnetic state to stabilize. Moreover, the magnitude of the
coercive field is too small and does not increase with the fall in
temperature. This suggests that there is no clustering of regions
having ferromagnetic character. It thus implies that the possibility
of segregation of magnetic entities forming regions of varied
stoichiometry should be ruled out.

M(H) at 10 K and 5 K show shifted loops with nearly zero coercive
magnetization at H = 0. Typically, pinching of the hysteresis loop
of this nature taking place along the magnetic field axis has been
observed in systems like small coated particles, inhomogeneous
materials, thin films and bilayer \cite{nog}. This effect is usually
observed when the FM/AFM system is either zero field cooled in a
demagnetized state or field cooled from above the Neel temperature
of the antiferromagnet \cite{adv}. It is related to a FM/AFM
exchange coupling interaction across the interface and believed to
be due to the formation and pinning of domains either in the FM or
in the AFM \cite{zha,mil}. The observation of such a feature in the
low temperature magnetization for the present sample suggests a
canted spin structure with ferromagnetic and antiferromagnetic spin
components in zero magnetic field, and the ferromagnetism is
field-induced. Hence the reduction in magnetization to almost zero
upon martensitic transformation and the T$^*$ signature is
apparently due to incipient antiferromagnetic coupling inherent in
the unit cell of the compound. The two interactions compete with
each other throughout the measurement temperature range giving rise
to complex behaviour in the magnetic properties of the Ni-Mn-In
system.

\begin{figure}
\centering \epsfig{file=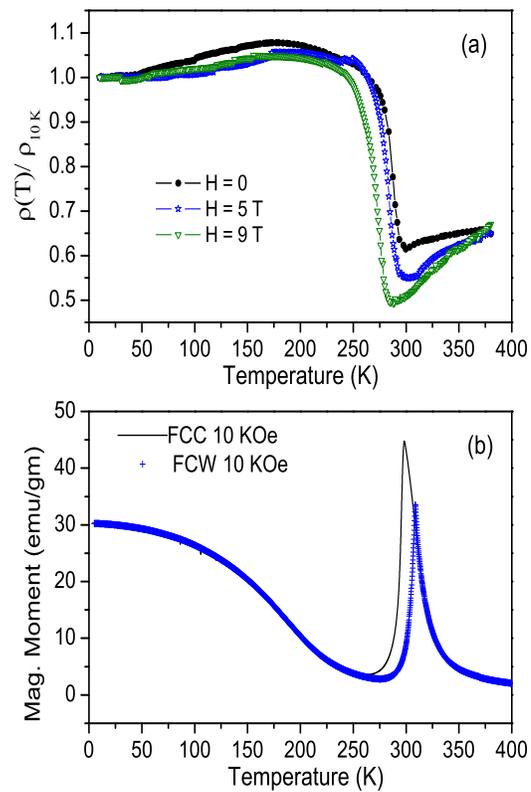, width=8cm, height=12cm}
\caption{\label{res}(Colour online)(a) Normalized resistivity
measured at different magnetic fields in the temperature range of 10
K to 380 K. (b) Temperature dependent magnetization measured in the
magnetic field of 10 KOe. The data was collected during the field
cooled state (FCC) and the subsequent warming (FCW) in the same
field.}
\end{figure}

Figure \ref{res}(a) represents the temperature dependence of
resistivity measured in the region 5 K to 380 K. The most striking
feature here is the large jump in the resistivity of
Ni$_{50}$Mn$_{35}$In$_{15}$ at room temperature. When viewed from
the high temperature side, a change in slope of $\rho(T)$ is seen at
about $\sim$ 310 K. This is the signature of ferromagnetic ordering.
At about 302 K, the resistivity sharply increases resulting in a
large step-like feature at the start of the martensitic transition.
With further fall in temperature the overall resistivity does not
seem to change much down to the lowest measurement temperature.

A small step or kink in resistivity has routinely been observed for
several other martensitic materials. Infact, such a feature along
with the associated thermal hysteresis has traditionally been
considered as the signature for martensitic transition (see for
example reference \cite{vasil}). The reason for the step /kink is
believed to be the trapping of electrons in the nested regions of
the Fermi surface due to long range structural ordering formed as a
consequence of martensitic phase change. And the nesting vector
corresponds to the modulation of martensite formed \cite{zhao, wil,
veli}. However, after the transformation is complete, $\rho$ in the
martensitic phase can be extrapolated to match with the curve
obtained before the transition had occurred. What makes the observed
step like feature special in Ni$_{50}$Mn$_{35}$In$_{15}$ is the
accompanying change in magnitude of resistivity. It may be noted
that the change in $\rho(T)$ is about $\sim$40\% in this case.

The anomalous feature in the $\rho(T)$ of
Ni$_{50}$Mn$_{35}$In$_{15}$ resembles to that observed in the
intermetallic compounds undergoing a transition to AFM state. A
typical example being CeFe$_2$ and its substitutional derivatives
\cite{N-ali, garde}. In such cases, a rise in resistivity after the
AFM transition is said to be due to the formation of super-zone gap.
Due to the establishment of the AFM sublattice, the underlying zone
boundaries get re-defined giving rise to a gap at the Fermi level.
The conduction electrons thus have to overcome this gap resulting in
a large anomaly in the transport property of the AFM state. The
magnetic properties of Ni$_{50}$Mn$_{35}$In$_{15}$ already reveals a
possibility of AFM interactions being present in the system along
with the underlying FM ones. Thus the anomalous feature in the
$\rho(T)$ and all other aforementioned measurements indicate that
the AFM interactions develops when the system heads towards the
structural transition. To ascertain this further, we measured
resistivity as a function of temperature in constant magnetic
fields. The $\rho(T)$ curves in the presence of different magnetic
fields show a very similar behavior to that at zero field. However,
the martensitic transition temperature shifts to lower value with
the increase in the magnetic field. The $\rho(T)$ curves at a
magnetic field of 5 T and 9 T are also shown in figure \ref{res}(a).
This trend implies that the magnetic field suppresses the structural
phase transition temperature and stabilizes the FM phase. This
observation is in congruence to that observed in figure \ref{res}(b)
where a decrease in the T$_M$ is observed in the field cooled data
recorded at 1 T field. Also, the hysteresis observed in the field
cooled and field warmed magnetization data further indicates the
first order nature of the martensitic transformation. The
antiferromagnetic interaction is seen to couple strongly with the
martensitic phase transformation.

Magnetism in Heusler alloys has always been fascinating and
continues to attract a lot of research interests till date
\cite{saso}. As follows from the magnetic properties of
Ni$_{50}$Mn$_{35}$In$_{15}$, the system exhibits complex interplay
between ferro- and antiferro- magnetic order. The competition
between these two magnetic interactions exists through the entire
temperature range of measurement up to T$_C$. Microscopically, the
formation of the competing phases can be related to the interatomic
exchange interactions that are dominated by separation between atoms
and the change in conduction electron density. In the Mn- based
Heusler systems, the spatial separation between the neighbouring Mn
atoms being comparatively large ($\sim$ 4\AA), a considerable direct
overlap of Mn {\it 3d} states is not observed here. Consequently, an
indirect RKKY type exchange mediated via the conduction electrons of
the system is often revoked to describe the magnetic ordering in
these materials \cite{kluber}. In addition, if such systems undergo
a martensitic transition, the change in interatomic distances upon
transformation are expected to highly manipulate the magnetic
interactions.

Amongst the stoichiometric Ni$_2$MnZ (Z = Ga, In, Sn, Sb), only
Ni$_2$MnGa undergoes a martensitic transformation. Experimentally,
all of them are ferromagnetic and have similar values of the Curie
temperature. In the case of Ni$_2$MnIn, the ferromagnetic ordering
takes place at T$_C$ = 315 \cite{web-book} and the martensitic
transformation is observed only in the non-stoichiometric, Mn rich
compositions. The ordered crystal structure in such Mn rich
composition contains excess Mn atoms at the In site in addition to
its regular ($1\over4$,$1\over4$,$1\over4$) sites. We have
previously studied the local crystal structure of these systems in
the cubic and martensitic phase and obtained the exact interatomic
separation between constituent atoms in both the phases
\cite{pab-jpd}. In the cubic phase, Ni$_{50}$Mn$_{35}$In$_{15}$ has
a lattice parameters of $\sim$ 6.04 \AA~. Accordingly, the Mn atoms
which substitute In atoms develop an additional Mn-Mn interaction at
$\sim$ 2.911 \AA~ with a 4.2 coordination number. While the
separation between Mn atoms present at its own site is 4.27\AA~ with
coordination number 12. Since the magnetic coupling between atoms is
governed by the interatomic separation, the Mn-Mn interactions at
4.27\AA~ is FM in nature. While the coupling between Mn atoms at
$\sim$ 2.91 \AA~ apart, must be AFM. The AFM nature of such
correlations have previously been anticipated from high resolution
neutron powder diffraction measurements on similar composition,
Ni$_{50}$Mn$_{34}$Sn$_{16}$ \cite{brown-sn}. However it is important
to note that no additional diffraction peaks corresponding to an
antiferromagnetic sublattice was observed in this measurements.

Moreover, it is also known from the local structure study of
Ni$_{50}$Mn$_{35}$In$_{15}$ that the cubic structure gets highly
unstable with Mn atoms moving with largest amplitude of displacement
as the T$_M$ is approached. Such vibration of Mn atoms from its
crystallographic position weakens the FM sublattice leading to its
collapse at the structural phase transformation. The subsequent
change in interatomic separation further uncovers the AFM sublattice
interactions. These AFM interactions drive the system to nearly zero
magnetic moment.

Once the structural transformation is complete and the martensitic
phase is fully established, the constituent atoms cease to move
vigorously. This is reflected in the low temperature bond distance
and the associated thermal mean square variation. The change in
crystal symmetry generates Mn-Mn bond at 2.89 \AA~ with coordination
number 4.2, while the Mn-Mn bonds at 4.27 \AA~ in the cubic phase
split into two correlations: 4.19 \AA~ with coordination number 8
and 4.4\AA~ with coordination number 4. The underlying magnetic
interactions also get re-established with such a change and the
associated FM/AFM interactions start competing resulting in an
anomaly like at T$^*$ in the temperature dependent magnetization
measurements. The relative strength of the two magnetic interactions
depends on the magnitude of the applied field. The average Mn-Mn
distance for ferromagnetic interaction (i.e. 4.19 \AA~ and 4.4\AA~)
in the martensitic phase equals the Mn-Mn distance in the cubic
phase. Thus though AFM interactions intensify during the process of
martensitic transition, the FM phase emerges upon completion of the
structural phase transformation and continues to dominate. Small
magnetic field values are sufficient to strengthen the FM order
further and help restore it. The two magnetic interactions compete
for dominance and continue to co-exist.

\section{Conclusion}
In conclusion, for the non-stoichiometric Ni-Mn-In compounds the
substituent has a crucial bearing on its magnetic properties. Based
on our ac susceptibility and dc magnetization study it is clear that
the origin of the anomalous magnetization behaviour of
Ni$_{50}$Mn$_{35}$In$_{15}$ and the exotic properties associated
with it are an outcome of competing FM/AFM magnetic interactions.
The AFM interactions manifests when the system heads towards
structural instability. In the low temperature region, the
short-range antiferromagnetic correlations are easily suppressed by
magnetic fields exceeding few hundred Oesterds once the structural
transformation is complete.

\begin{figure}
\centering \epsfig{file=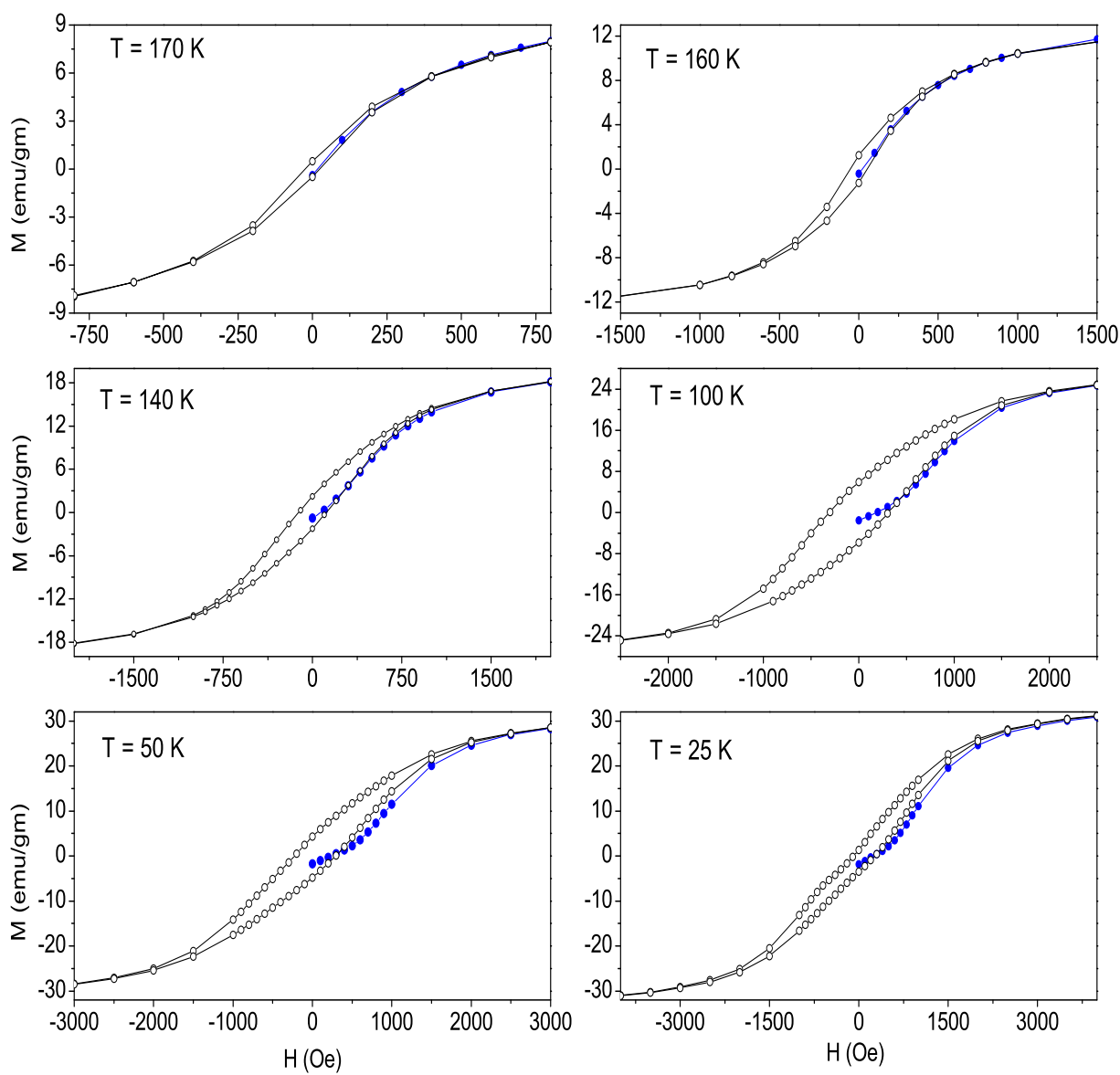, width=18cm, height=18cm}
\caption{\label{all-loop}(Colour online) M(H) loops at different
temperatures below T$^*$. Measurements were carried out by sweeping
the field from +5 T to -5 T. A magnified view for a smaller field
interval is shown here for clarity. Data points for the virgin curve
is shown as solid circles while for the rest of the loop it is shown
with empty circles.}
\end{figure}

\begin{figure}
\centering \epsfig{file=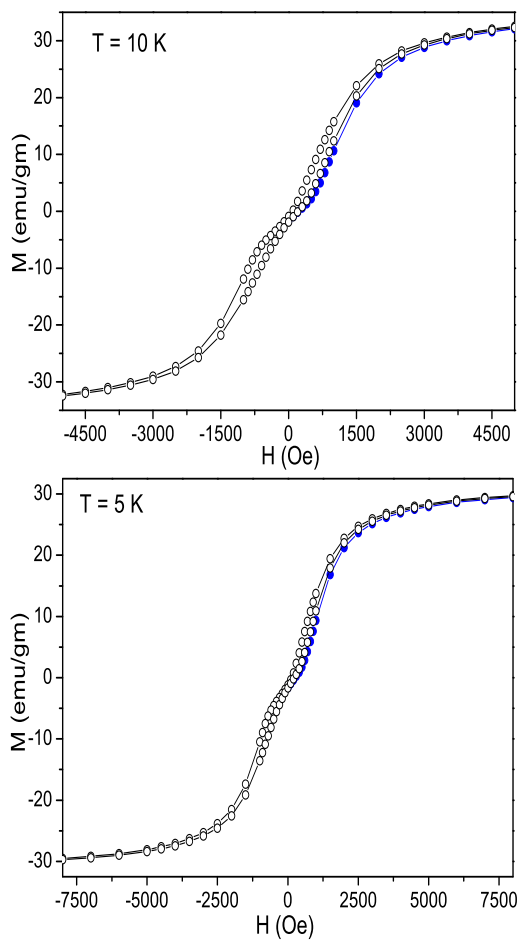, width=8cm, height=14cm}
\caption{\label{all-loop}(Colour online) The double shifted loops
obtained at 10 K and 5K are represented. The virgin curve is shown
as solid circles while for the rest of the loop it is shown with
empty circles. Measurements were carried out from +5T to -5T. A
magnified view is shown here.}
\end{figure}

\end{document}